\documentstyle[preprint,aps]{revtex}
\newcommand{\beq}{\begin{equation}}
\newcommand{\eeq}{\end{equation}}
\newcommand{\beqar}{\begin{eqnarray}}
\newcommand{\eeqar}{\end{eqnarray}}
\newcommand{\sect}[2]{\vspace*{0.5\baselineskip}\hspace*{-\parindent}{\bf
#1.}~{\bf
#2}\hspace*{\parindent}}

\tighten
\begin{document}
\draft
\preprint{\parbox[t]{85mm}{Preprint Numbers: \parbox[t]{50mm}
        {ANL-PHY-7718-TH-94\\ KSUCNR-004-94}
}}
\title{Charge symmetry breaking via \mbox{\boldmath $\rho$}-\mbox{\boldmath
$\omega$} mixing \\from model quark-gluon dynamics}
\author{K. L. Mitchell\footnotemark[1], P. C. Tandy\footnotemark[1],
C. D. Roberts\footnotemark[2] and R. T. Cahill\footnotemark[3]}
\address{\footnotemark[1]
Center for Nuclear Research, Department of Physics, Kent State University,
Kent, Ohio 44242\\
\footnotemark[2]
Physics Division, Argonne National Laboratory, Argonne, Illinois 60439\\
\footnotemark[3]
School of Physical Sciences, Flinders University of South Australia,
SA 5042, Australia}
\date{\today}
\maketitle
\begin{abstract}
The quark-loop contribution to the $\rho^0-\omega$ mixing self-energy
function is calculated using a phenomenologically successful QCD-based model
field theory in which the $\rho^0$ and $\omega$ mesons are composite
$\bar{q}q$ bound states.  In this calculation the dressed quark propagator,
obtained from a model Dyson-Schwinger equation, is confining. In contrast to
previous studies, the meson-$\bar{q}q$ vertex functions are
characterised by a strength and range determined by the dynamics of the
model; and the calculated off-mass-shell behaviour of the mixing amplitude
includes the contribution from the calculated diagonal meson self-energies.
The mixing amplitude is shown to be very sensitive to the small isovector
component of dynamical chiral symmetry breaking.  The spacelike quark-loop
mixing-amplitude generates an insignificant charge symmetry breaking nuclear
force.
\end{abstract}
\hspace*{-\parindent}{\bf 1. Introduction.}\hspace*{\parindent}
The on-mass-shell mixing of $\rho^{0}$ and $\omega$ mesons can be represented
via a $2 \times 2$ matrix, ${\cal M}^2$, with diagonal elements $m_{\rho}^2$
and $m_{\omega}^2$ and both off-diagonal elements equal to the mixing
amplitude $\Pi$.  The established value of $\Pi$, deduced from $e^{+} e^{-}
\rightarrow 2\pi$ data, is $-4520 \pm 600$~MeV$^2$\cite{coon}.
Recently~\cite{CS94} the systematics of electromagnetic and semi-strong mass
splittings of the hadrons have been used to deduce a universal strength of
$-5000$~MeV$^2$ for isovector mixing; i.e., for both $\rho^{0}-\omega$ and
$\pi-\eta$.  The contributions that this mixing makes to the various charge
symmetry breaking (CSB) phenomena in nuclear physics have been much
studied\cite{coon,review}. The central element is the corresponding CSB
contribution to the nucleon-nucleon potential, which is obtained in the boson
exchange approach by using
\begin{equation}
\frac{-\Pi}{(k^2-m_{\rho}^2)\,(k^2-m_{\omega}^2)}
\end{equation}
as the off-diagonal propagator.  Without evidence of significant momentum
dependence, $\Pi$ has been taken as a constant in nuclear CSB applications
and the resulting phenomenology has been very successful.~\cite{coon,review}
In a number of circumstances, the $\rho^{0}- \omega$ CSB nuclear potential
thus obtained has been seen to dominate over other mechanisms, such as
electromagnetic effects, the mass difference of neutrons and protons and
$\pi^{0}-\eta$ mixing.  A particularly striking example is the IUCF
measurement\cite{IUCFnp} of the analyzing power difference in $np$ scattering
at $180$~MeV.

In QCD, the mixing arises because of the small $u$-$d$ current-quark-mass
difference, which induces a difference between the propagation of the
$\bar{u}u$ and $\bar{d}d$ components, and one expects a momentum dependent
off-diagonal self-energy.  Knowledge of the off-mass-shell behaviour is
needed to extrapolate to spacelike momenta appropriate to meson exchange
between nucleons.  The simplest possible contribution to $\Pi(k^2)$ is a
single quark loop and the initial investigation employed free constituent
quark propagators and phenomenological meson-$q$-$\overline{q}$-vertex form
factors to obtain a closed form expression\cite{GHT}. The relevance of the
result is clouded somewhat by the presence of a spurious quark production
threshold, inconsistent with the notion of quark confinement.  To avoid this
unphysical feature, which arises near the meson mass-shell, a subsequent
work\cite{Krein} employed a simple, phenomenological form of dynamical quark
propagator that implements confinement through the absence of a mass-shell
pole.  Nevertheless the central result remained the same: for spacelike
momenta appropriate to vector exchange between nucleons, the mixing is so
much weaker than the mass-shell value that the corresponding CSB nuclear
potential is negligible.  A recent QCD sum rule investigation\cite{qcdsum}
found a somewhat stronger momentum dependence, with $\Pi(k^2)$ reaching
the negative of the mass-shell value at $k^2\approx 0$.
In contrast, the momentum dependence of the $\pi^{0}-\eta$ mixing amplitude
calculated from chiral perturbation theory appears to be considerably
weaker~\cite{MG}.

There are a number of issues that suggest the quark loop calculations may not
yet be reliable for this process.  An important consideration for
$\rho^{0}-\omega$ mixing is that it is driven by the small isovector
component of the quark propagator, which in turn is driven by the isovector
current mass $m_u -m_d$. A direct parametrisation of a propagator model is
usually constrained by a fit to quantities related to dynamical chiral
symmetry breaking (DCSB), such as $f_{\pi}$, \mbox{$<\bar{q}q>$} and
$m_{\pi}$.  These are isoscalar constraints and are not sufficient here.  The
explicit chiral symmetry breaking (ECSB) generated by current masses is
weaker but its role is magnified in the isovector propagator and some
dynamical guidance is to be preferred.  The low momentum behaviour of the
dressed quark propagator can be calculated using a dynamical model
Dyson-Schwinger equation (DSE).  However, typical DSE studies are carried out
numerically in Euclidean space; i.e., at spacelike momenta, where maximal use
can be made of the constraints from perturbation theory and the
renormalisation group in QCD\cite{RW94,WKR}.
Such numerical studies do not provide readily accessible information about
the behaviour at real, time-like quark momenta nor the domain of complex
momenta sampled in a Euclidean treatment of meson propagators at time-like
meson momenta.

The above issues are addressed in this work through use of a recently
developed solution\cite{entire} of a model DSE that yields the quark
propagator in closed form as an entire function in the complex momentum
plane.  This structure can be interpreted as representing confined
quarks\cite{RW94,RWK92} and provides an unambiguous representation over the
complex momentum domain required by the quark loop. The dependence upon the
current mass is explicit in the model and is produced by the DSE dynamics.

We calculate the mixed $\rho^{0}-\omega$ self-energy function produced by the
vector meson sector of a QCD-based model field theory\cite{CR85,GCM}, the
Global Colour-symmetry Model (GCM) [for a review see ~\cite{rtc}], which
formalises the coupled Dyson-Schwinger--Bethe-Salpeter equation approach to
QCD phenomenology\cite{RW94} and allows several internally consistent
features to be implemented for the first time.  The off-mass-shell behaviour
we obtain for the mixing amplitude includes the contribution from the
off-mass-shell structure of the composite $\rho$ and $\omega$ propagators. In
addition, the strength and range parameters used to model the
meson-\mbox{$\bar{q}q$} Bethe-Salpeter amplitudes are determined by the meson
mass-shell dynamics within the GCM.  This eliminates the need for several
assumptions made in previous studies.  A short account of some of this work
has recently been presented\cite{MT}.

\sect{2}{\mbox{\protect\boldmath $\rho^0$}-\mbox{\protect\boldmath $\omega$}
Mixing Amplitude.} The description of the $\rho^0$ and $\omega$ mesons as
$\bar{q}$-$q$ composites is obtained from a phenomenologically successful,
QCD-based model field theory\cite{CR85,GCM}, which is defined by the action:
\beq
S[\bar{q},q]=\int d^{4}x \,\bar{q}(x)  \,
\bigl( \gamma \cdot \partial _{x} + M \bigr) \,q(x)
+ \case{1}{2} \int d^{4}x d^{4}y\,
         j_{\mu }^{a}(x) g^{2} D(x-y)j_{\mu }^{a}(y),
\label{action}
\eeq
where $M$ is the current-quark-mass matrix,
$j_{\mu}^{a}(x)=\bar{q}(x)\frac{\lambda ^{a}}{2}\gamma _{\mu}q(x)$ is the
quark current, and $D(x-y)$ is a model effective-gluon-propagator in a
Feynman-like gauge.  Here we consider only $u$ and $d$ flavors and use a
Euclidean metric throughout such that \mbox{$a \cdot b= a_\mu b_\mu$} and
\mbox{$\{\gamma_\mu ,\gamma_\nu \}=2\delta_{\mu\nu}$}, with
$\gamma_\mu= \gamma_\mu^\dagger$.

After bosonisation and expansion about the classical vacuum [by which is
meant the minimum of the action], the tree-level effective action for the
mesons is\cite{GCMmeson}
\beqar
\hat{S} = Tr \sum_{n=2}^{\infty } & & \frac{(-1)^{n}}{n}
\bigl[
          G_{0}( i\gamma_{5}\vec{\tau}\cdot \vec{\pi}
                +i\gamma_{\nu} \omega_{\nu}
                +i\gamma_{\nu} \vec{\tau} \cdot \vec{\rho}_{\nu}
                + \cdot \cdot \cdot )
\bigr] ^{n} \nonumber \\
&+& 9 \int d^{4}xd^{4}y
\frac{    \frac{1}{2}\vec{\pi} \cdot \vec{\pi}
          + \omega^2 + \vec{\rho} \cdot \vec{\rho}+ \cdot \cdot \cdot }
{2g^2D(x-y)}  ,
\label{meson}
\eeqar
where the meson fields are bilocal fields corresponding to bilocal combinations
of quark fields; for example, \mbox{$\omega_{\mu}(x,y) \sim \bar{q}(y)i
\gamma_{\mu}q(x)$}.  The dressed quark propagator, $G_0$, that appears here has
a self-energy determined by the classical vacuum configurations of the bilocal
fields.  We use a momentum representation in which $q$ is conjugate to $x-y$
and
$P$ is conjugate to $(x+y)/2$.  The vector bilocal fields are represented by,
for
example, $\gamma_{\mu}\omega_{\mu}(q;P)= \Gamma(q;P)\gamma_{\mu} T_{\mu \nu}(P)
\omega_{\nu}(P)$, where \mbox{$T_{\mu \nu}(P)= \delta_{\mu \nu} -P_{\mu}
P_{\nu}/P^2$} and $\Gamma(q;P)$, a Lorentz-scalar function, is the dominant
Bethe-Salpeter amplitude in the vector $\bar{q}q$
channel\cite{GCMmeson,JM93}.

We are interested in tree-level coupling in the $\rho^0 - \omega$ sector.  In
a matrix notation where $V_{\mu}$ denotes the column of transverse fields
$(\vec{\rho}_{\mu}, \omega)$, the GCM action up to second order in the fields
is
\beq
\hat{S}[\rho,\omega] = \frac{1}{2} \int \frac{d^{4}P}{(2\pi)^4}
V^{T}_{\mu}(-P) \Bigl[ \Delta^{-1}_{\mu \nu}(P) + \Pi_{\mu \nu}(P) \Bigr]
V_{\nu}(P) ,
\label{romaction}
\eeq
where $\Delta^{-1}$ is diagonal and $\Pi$ is the off-diagonal $\rho^0 -
\omega$ self-energy.  The diagonal inverse propagator is
\beq
\Delta^{-1}_{\mu \nu}(P) = \int \frac{d^4q}{(2\pi)^4}
tr\left[ G_{0}(q_{-})i\gamma_{\mu} f G_{0}(q_{+})i\gamma_{\nu}f\right]
\Gamma^2(q;P)
+ 9 \delta_{\mu \nu} \int d^4 r \frac{ \Gamma^2(r;P) }{g^2D(r)},
\label{delinv}
\eeq
with $tr$ denoting a trace over spin, flavor and color.  Here $f$ are the
relevant flavor factors [$1$ for the $\omega$, and $\vec{\tau}$ for the
$\vec{\rho}$] and \mbox{$q_{\pm}=q \pm P/2$}, where $P$ is the meson momentum
and $q$ is the loop momentum associated with the $\bar{q}q$ substructure.
The mixed self-energy, $\Pi_{\mu \nu}(P)$, is given by an expression
identical to the first term of (\ref{delinv}) except that one of the flavor
factors is $\tau_{3}$ and the other is $1$, thus it is nonzero only if
$G_{0}$ has an isovector component.

At tree level the diagonal terms are the same for $\rho^0$ and $\omega$ and
the general form of the transverse component is
\begin{equation}
\Delta^{-1}_{\mu\nu}(P^2) = T_{\mu \nu}(P) \, (P^2+M_{V}^2) \, Z(P^2),
\end{equation}
where the position of the zero defines the physical mass $M_{V}$.  The meson
fields may be rescaled so that at least the on-mass-shell value of
$\sqrt{Z}$ is absorbed into the fields to produce a physical normalization.
We elect
to absorb $\sqrt{Z(P^2)}$ into the fields so that the diagonal propagators
have the standard form characterised only by the physical mass. The
corresponding mixed self-energy $\Pi_{\mu \nu}(P)/Z(P^2)$
then contains the off-mass-shell structure
appropriate for use in meson exchange applications that use the standard
point-meson form for diagonal propagators. The net result of this is that the
renormalised version of (\ref{delinv}) is obtained via the replacement:
\begin{equation}
\Gamma(q;P) \rightarrow F(q;P)=\frac{\Gamma(q;P)}{\sqrt{Z(P^2)}}.
\label{F}
\end{equation}
The resulting vertex function, $F(q;P)$, contains the strength of the
meson-$\bar{q}q$ coupling as determined by the model.  On the mass-shell
this is equivalent to imposing the standard physical normalisation for a
Bethe-Salpeter amplitude\cite{IZ}.

The corresponding mixed self-energy
\beq
\Pi_{\mu \nu}(P) = \int \frac{d^4q}{(2\pi)^4}
tr\left[ G_{0}(q_{-})i\gamma_{\mu} \tau_{3} G_{0}(q_{+})i\gamma_{\nu}\right]
F^2(q;P)
\label{pimn}
\eeq
has a transverse component, \mbox{$\Pi_{\mu \nu}(P)=T_{\mu \nu}(P)
\Pi_{T}(P^2)$}, which yields the mixing amplitude as the following difference
of quark loop integrals:
\begin{equation}
\label{Piud}
\Pi_{T}(P^2) = \Pi^u_{T}(P^2) - \Pi^d_{T}(P^2).
\end{equation}
With the quark propagator in the form
\mbox{$G_{0}(p)=-i\gamma \cdot p \sigma_{V}(p^2) + \sigma_{s}(p^2)$}, the
required quark loop integrals are
\begin{eqnarray}
\lefteqn{\Pi^i_{T}(P^2) =} \label{pi} \\
& &  - \case{12}{(2\pi)^3} \int^{\infty}_{0} ds s
F^2(s;P^2) \int^{+1}_{-1} du (1-u^2)^{ \frac{1}{2} }
\Bigl[ \bigl( \case{s}{3} (1+2u^2) - \case{1}{4}P^2 \bigr)
       \sigma^{i}_{v}(+) \sigma^{i}_{v}(-)
      +\sigma^{i}_{s}(+) \sigma^{i}_{s}(-) \Bigr], \nonumber
\eeqar
where $s=q^2$, \mbox{$u=\hat{q} \cdot \hat{P}$} and $\sigma(\pm)$ denotes
$\sigma(q_{\pm}^2)$, where \mbox{$q_{\pm}^2 = s + P^2/4 \pm u\sqrt{s P^2}$}.
For time-like meson momentum, where $P^2 < 0$, \mbox{$\sqrt{P^2} \rightarrow
i \sqrt{-P^2}$} and $\sigma(-) = \sigma^{\star}(+)$. By assumption the vertex
function, $F$, is independent of the angle $u$.

\sect{3}{Quark propagator.}
Because the momentum region of interest [\mbox{$P^2 \geq -m_{\omega}^2$}]
includes a time-like sector, the loop integral (\ref{pi}) requires knowledge
of the propagator amplitudes $\sigma_{s}(z)$ and $\sigma_{v}(z)$ in the
domain of the complex $z=x+iy$ plane enclosed by the parabola
\mbox{$|y|=m_{\omega}\sqrt{x+m_{\omega}^2/4}$} with
\mbox{$x \geq -m_{\omega}^2/4$}.   A fit to soft chiral quantities [such as
$\langle\bar{q}q\rangle$, $f_{\pi}$, $m_{\pi}$, $r_{\pi}$] does not probe a
propagator model over this domain.  [The corresponding quark loop appropriate
for pion physics also requires knowledge of the quark propagator in the
complex plane but in a region whose area is reduced by a factor of $(
m_{\omega} / m_{\pi} )^2 \approx 25$ and hence the chiral limit
($m_{\pi}\rightarrow 0$) captures the dominant physics.]

The quark propagator given by Ref.~\cite{entire} provides a behaviour in the
complex domain that is accountable to DSE dynamics.  It is conveniently
expressed in terms of dimensionless amplitudes
\mbox{$\bar{\sigma}_{s}(x) = \lambda \sigma_{s}(q^2)$} and
\mbox{$\bar{\sigma}_{v}(x) = \lambda^2 \sigma_{v}(q^2)$}, where
$x=q^2/ \lambda^2$ and $\lambda$ is the momentum scale. The amplitudes
$\bar{\sigma}$ are also explicit functions of the dimensionless current-quark
mass $\bar{m}= m/ \lambda$. This DSE model employs a momentum-space delta
function for the gluon propagator, with a strength set by $\lambda^2$, and a
dressed quark-gluon vertex\cite{CP}, which is compatible with the
Ward-Takahashi identity.  The amplitudes $\bar{\sigma}$ are entire functions
in the complex momentum plane with an essential singularity at timelike
infinity.  The exact expressions are a little complicated when the current
mass is included.  We have verified that for typical current masses [$\sim 10
$~MeV], the following simple forms
\beq
\bar{\sigma}_{s}(x)= c(\bar{m}) e^{-2x} +
\frac{ \bar{m} }{x}(1-e^{-2x}) ,
\label{ssb}
\eeq
\beq
\bar{\sigma}_{v}(x) = \frac{ e^{-2x}-(1-2x) }{2x^2} -
\bar{m}~c(\bar{m}) e^{-2x} ,
\label{svb}
\eeq
are accurate representations [to order $10^{-4}$] on the domain of complex
momenta encountered in the quark loop.

These forms also accurately represent the leading behaviour of the quark
propagator at large spacelike $q^2$ in the model and in QCD.  The constant
$c(\bar{m})$, which represents the strength of DCSB, is not determined by the
model DSE solution; the reason being that the model of Ref.~\cite{entire} has
only two scales [momentum $\lambda$ and mass $m$].  A third scale is
necessary to fix $c(\bar{m})$.  In a more elaborate model, such as that of
Ref.~\cite{WKR}, this is provided by $\Lambda_{QCD}$, which is introduced by
the inclusion of the known large spacelike-$q^2$ behaviour of the gluon
propagator.  However, a numerical solution is then necessary and the
extension into the complex momentum plane is not unique.  We consider
(\ref{ssb}) and (\ref{svb}), with $c(\bar{m})$ a parameter to be determined
below, to provide a realistic model propagator for a confined quark in QCD
over the region of the complex plane sampled in (\ref{pi}).

To fix the values of the parameters $\lambda$ and $c_0= [c_u + c_d]/2$ we
follow Ref.~\cite{pipi} and use (\ref{ssb}) and (\ref{svb}) to fit
$\langle\overline{q}q\rangle$, $f_\pi$, $r_\pi$ and the $\pi$-$\pi$
scattering lengths: $a_0^0$, $a_0^2$, $a_1^1$, $a_2^0$.  With \mbox{$(m_u
+m_d)/2= 7.5$ MeV} this leads to best fit values of $\lambda = 0.893$~GeV and
$c_0=0.523$, which yield
\mbox{$\langle\bar{u}u\rangle=\langle\bar{d}d\rangle= (183 \;{\rm MeV})^3$},
\mbox{$f_{\pi}=90$ MeV}, $r_\pi = 0.56$~fm and a mean deviation of $19\%$
from the experimental $\pi\pi$ scattering lengths.

\sect{4}{Bethe-Salpeter amplitude.}
The dominant vector meson Bethe-Salpeter amplitude, $F(q;P)$, also appears in
(\ref{pi}).  This may be calculated in the GCM using a variational form of
the off-mass-shell generalised-ladder-approximation Bethe-Salpeter eigenvalue
equation ~\cite{rtc}, and a simple, reasonable approximation for the shape on
the mass-shell is provided by the $P$-independent form $\Gamma(q;P)=
\exp(-q^2/a^2)$, with $a=0.62$~GeV. To simplify the computations here  the
renormalisation function, $Z(P^2)$, was obtained from (\ref{delinv}) after the
second [constant] term is set to produce the vector mass $783$~~MeV.  The
normalised amplitude $F(q;P)$ is then obtained from (\ref{F}).

\sect{5}{Isovector component of the quark propagator and
\mbox{\boldmath $\rho^0$}-\mbox{\boldmath $\omega$} mixing.}
It only remains now to set the isovector component of the quark propagator,
which is a measure of the isovector character of DCSB.  With the
representation
\mbox{$G_{0}^{-1}(q^2)= i\gamma \cdot q A(m,q^2)+B(m,q^2)$}, one commonly
used assumption is that \mbox{$A(m,q^2)
\approx A(0,q^2)$} and \mbox{$B(m,q^2) \approx B(0,q^2)+m$}.  This is the
procedure followed in the one other calculation of $\rho^0-\omega$ mixing
that used a dynamical quark propagator\cite{Krein}.  It is convenient because
only the chiral limit amplitudes need to be modelled.  With the
$\overline{m}=0$ forms of $A$ and $B$ extracted from (\ref{ssb}) and
(\ref{svb}), this yields the prediction for the $\rho^0-\omega$ mixing
amplitude shown as the dotted line in Fig.~1.  [Throughout this work we use
\mbox{$m_d - m_u = 5$~MeV}.]

The momentum dependence obtained is quite similar to that found in earlier
work\cite{GHT,Krein}.  The experimental mass-shell mixing strength was
produced in the work of Ref.~\cite{Krein} with use of phenomenological
meson-quark coupling strengths \mbox{$g_{\rho}=4$} and \mbox{$g_{\omega}=5$}
deduced by simple scaling arguments from empirical meson-nucleon coupling
constants.  In the present approach, however, the vertex strength is an
outcome of the model, and the comparable measure from (\ref{F}) is
\mbox{$g_{\rho}=g_{\omega}=F(q=0;P^2=-M_{V}^2)$} which produces the
larger value $16.4$.  Nevertheless, we obtain only $50\%$ of the experimental
mixing value.  The quark vertex alone is not an accurate guide to the net
coupling strength between hadrons because it does not account for the
particular way the hadron dynamics samples the quark propagator amplitudes,
especially the wave function renormalisation function.  We feel the
mass-shell condition used here provides the internally consistent vertex
normalisation for the meson model.

The above estimate for the isovector propagator can be improved through the
use of solutions from a realistic model DSE to directly determine the
isovector component of $c(\overline{m})$ in (\ref{ssb}) and (\ref{svb}).  [We
note that in realistic DSE studies $A(m,0)$ decreases with increasing $m$ and
$B(m,0)$ is badly underestimated by $B(0,0)+m$.]  From the model of
Ref.~\cite{WKR}, which uses a model gluon propagator that is the sum of a
momentum-space delta function and the one-loop renormalisation group result
together with a dressed quark-gluon vertex, one infers a value of
\mbox{$c_1 = (c_u -c_d)/2 = 0.006$} for $m_u=5$~MeV and $m_d=10$~MeV [by
matching $c(\overline{m})$ to the $\overline{m}$ dependence of the solutions
at $q^2=0$ and using the $C=500$, $\tau = e$, $N_f=4$ results from
Ref.~\cite{WKR}].  This is an indication of the size of the isovector
component of DCSB, which is seen to be a $1\%$ effect in this case.

The mixing amplitude obtained with this parameter set is given by the dashed
curve in Fig.~1: less than $25\%$ of the mass-shell mixing strength is
accounted for.  Of course, the above arguments are simply a guide. If $c_1$
is treated as a free parameter a fit to the mass-shell mixing amplitude
requires \mbox{$c_1 = 0.009$}, which yields the momentum dependence shown by
the solid line in Fig.~1.  Clearly, the quark loop mixing mechanism is
extremely sensitive to the small isovector component of DCSB.

Qualitatively, the momentum dependence of the mixing amplitude is insensitive
to the form of the model Bethe-Salpeter amplitudes at the vertices. The
insert in Fig.~1 illustrates this by comparing the previously mentioned
result [solid line] from the simple gaussian amplitude to the result
[dash-dot line] obtained using the dominant $\rho$-amplitude obtained in a
recent Bethe-Salpeter calculation~\cite{JM93}.  Both representations for the
on-mass-shell meson vertex give essentially identical results.

In Fig. 2 we isolate the part of the off-mass-shell behaviour of the mixing
amplitude that is produced by the momentum dependence of the function
$Z(P^2)$.  This is a measure of the extent to which the diagonal meson
propagators depart from the standard point form due to their
\mbox{$\bar{q}q$} substructure.  The solid curve is the result previously
shown and includes the off-shell contributions from the propagators, while
the dashed curve has $Z(P^2)$ held fixed at the mass-shell value to remove
them.  Clearly this contribution is significant for \mbox{$P^2>1~$GeV$^2$}.

\sect{6}{CSB potential.}
In Fig.~3 we display the CSB NN potential $V_{\rho \omega}(r)$ calculated
from the standard momentum space form\cite{coon}
\beq
V_{\rho \omega}(P) = - \frac{
g_{\rho N}\, g_{\omega N}\,
        F_{\rho N}(P^2)\, \Pi_{T}(P^2) \,F_{\omega N}(P^2)}
        { (P^2 + m_{\rho}^2) (P^2 + m_{\omega}^2) }~ .
\label{V}
\eeq
In accordance with the Bonn meson exchange NN interaction model\cite{bonn},
monopole form-factors for $\rho NN$ and $\omega NN$ coupling are used with a
common range of $1.5$~GeV, and the coupling constants are taken to be
\mbox{$g^2_{\rho N}/4\pi=0.41$} and \mbox{$g^2_{\omega N}/4\pi=10.6$}.  The
mixing amplitudes in Fig. 2 give rise to the potentials shown by the solid
and dotted curves in Fig. 3.  The small difference between them indicates
that the off-shell structure of the diagonal propagators of the composite
vector mesons does not make a signficant contribution to the potential.  This
is because $Z(P^2)$ departs significantly from its mass-shell value only for
momenta that are well suppressed by the meson-nucleon form factors.
Variations of $\pm 20\%$ in the form-factor range do not change this
conclusion.  The dashed curve is the result that follows from the mass shell
assumption, \mbox{$\Pi_{T}(P^2) \approx \Pi_{T}(-M_{V}^2)$}.

The typical quark loop result for the nuclear CSB potential is very much
weaker than the potential produced by the mass-shell assumption because the
mixing amplitude near \mbox{$P^2 \approx 0$} is about four times smaller than
the empirical mass-shell value.  This is a general consequence of a node at
very low momentum and an almost linear connection with the mass-shell
point.

There are uncertainties in QCD-models at this scale but our investigations
have shown no indications that the main conclusion above is affected.  For
example, the off-mass-shell behaviour of the meson-quark vertex is not
guaranteed to be represented well by a single mass-shell Bethe-Salpeter
amplitude.  In this regard, we have employed an estimate for the relevant
$P^2>0$ extrapolation of the gaussian representation of $\Gamma(q;P)$
described earlier. This can be defined within the GCM by the variational
approach\cite{rtc} to the eigenvalue problem based on (\ref{delinv}).
The dominant effect is incorporated by a momentum-dependent range $a(P^2)$
for the $q^2$ behaviour that increases with increasing spacelike
momentum. The net result is an increased spacelike suppression of the mixing
amplitude with a typically smaller value at \mbox{$P^2 \approx 0$}.

\sect{7}{Quark loop generates weak mixing.}
The quark loop mechanism will, in general, produce a weak mixing at small
spacelike momenta.  This can be seen from an analysis of the loop integral
in (\ref{Piud}) and (\ref{pi}) at first order in the isovector components of
$\sigma_{v}$ and $\sigma_{s}$ for $P^2>0$.

{}From (\ref{svb}), both the isoscalar and isovector $\sigma_v$ are always
positive. Thus the vector contribution to \mbox{$\Pi_{T}(P^2)$} is positive
for sufficiently large $P^2>0$.  From (\ref{ssb}), the isoscalar $\sigma_s$
is always positive, while the isovector component has a small positive
contribution from the first term, and a dominant negative contribution from
the second term. Thus the scalar contribution to \mbox{$\Pi_{T}(P^2)$} is
also positive. Since the vector contribution is dominant, there will be a
change in sign at small momenta driven mostly by the sign change in the
coefficient of the vector part of the integrand in (\ref{pi}). The position
of the node is approximately \mbox{$P_{0}^2 \approx <4s(1+2u^2)/3>$} with the
expectation value estimated from the dominant behaviour of the remaining
integrand in (\ref{pi}).  Using only the typical momentum scales that
characterise the propagator and the vertex, we obtain
\mbox{$P_{0}^2 \approx 0.5~$GeV$^2$} which is in accord with the
numerical results.

On the time-like side of the node, the linear factor of $P^2$ in (\ref{pi})
represents the behaviour well.

These observations hold equally well if a free constituent quark propagator
is used.

\sect{8}{Summary and conclusions.}
We have calculated the quark loop contribution to the $\rho^0$-$\omega$
mixing amplitude in a way that retains, for the first time, a number of
self-consistent features that follow from use of a QCD-based model field
theory. These include DSE model guidance for the quark propagator at the
time-like and complex momenta that arise, calculation of the strength
and range of the vertex function within the model, and the contribution
to the off-shell behaviour arising from the composite nature of the
diagonal meson propagators.

The mass-shell value of the mixing amplitude is found to be very sensitive to
the isovector component of DCSB.  For the behaviour at spacelike momenta
relevant to the CSB component of the NN interaction, none of the new elements
we include alter the conclusion that the quark loop mechanism alone generates
an insignificant CSB potential.

This failure suggests that other isospin-symmetry breaking intermediate
states, such as the virtual pion-loop process: \mbox{$\rho \rightarrow \pi\pi
\rightarrow \omega$} through a G-parity violating $\omega\pi\pi$ vertex,
may contribute significantly to $\Pi_{T}(P^2)$.  In this connection, and in
the context of the field theoretical model considered herein, the
$\omega\pi\pi$ vertex occurs as a quark loop which is non-zero for \mbox{$m_u
\neq m_d$} and a recent estimate\cite{MT} indicates that
\mbox{$g_{\omega \pi \pi}/g_{\rho \pi \pi} \approx
\Pi_{T}(-M^2_{V})/m_{\omega}^2$}.  It is therefore possible that the pion
loop mechanism could make a significant contribution to $\rho^0-\omega$
mixing and the associated NN CSB potential should be investigated.  This is
especially true since the imaginary part of such an amplitude relates to the
same interference effect in the $2\pi$ decay channel that is the basis of the
accepted experimental value of $\rho^0-\omega$ mixing.

\vspace*{0.5\baselineskip}\hspace*{-\parindent}{\bf
Acknowledgments}\hspace*{\parindent}
This work was supported in part by the National Science Foundation under
Grant Nos. PHY91-13117 and INT92-15223.  The work of CDR was also supported
by the US Department of Energy, Nuclear Physics Division, under contract
number W-31-109-ENG-38.  We acknowledge helpful discussions with S.A. Coon,
B.H.J. McKellar, A.W. Thomas and A.G. Williams.  CDR and PCT are grateful for
the hospitality extended to them by the School of Physical Sciences, Flinders
University of South Australia, and by the School of Physics, University of
Melbourne, during visits in April and July 1993.

\newpage
\begin{figure}
\caption{The quark loop contribution to the $\rho^0-\omega$ mixing amplitude.
Here $P^2>0$ represents spacelike momentum.  The solid curve is obtained with
the quark propagator parameter $c_1$ adjusted to fit the experimental mixing
value; the dashed curve uses the value inferred from a realistic DSE
solution; the dotted curve is produced by the common assumption that a
current mass provides a constant shift to the scalar self-energy function.
The insert demonstrates that the mass-shell vector Bethe-Salpeter amplitude
does not introduce significant model dependence.
\label{fig1}}
\end{figure}
\begin{figure}
\caption{The $\rho^0-\omega$ mixing amplitude with
(solid curve) and without (dashed curve) the off-mass-shell contribution
from the calculated diagonal propagators of the $\bar{q}q$ mesons.
    \label{fig2}}
\end{figure}
\begin{figure}
\caption{The $\rho^0-\omega$ mixing contribution to the CSB NN potential
form factor.  The dashed curve is obtained under the assumption that the
mixing amplitude is a constant given by the mass-shell value.  The quark loop
calculation is shown by the solid curve when the off-mass-shell contribution
from the diagonal meson propagators is included, and by the dotted curve when
it is not.  \label{fig3}}
\end{figure}
\end{document}